# Visual Discovery of Fading Events in the Wolf-Rayet Star WR 80

Rod Stubbings

*Tetoora Road Observatory, 2643 Warragul-Korumburra Road, Tetoora Road, VIC 3821, Australia*

*stubbo@dcsi.net.au*

**Abstract**

Long-term visual and CCD *V*-band monitoring of the heavily reddened Wolf-Rayet star WR 80 from 2017–2026 reveals optical fading events that depart from its long-term stable baseline. On 26 April 2017 (JD 2457870), I discovered a deep minimum using a 22-inch (0.56-m) telescope, with the source plunging from a baseline of magnitude 14.6 down to magnitude 15.8. This 6-day event was subsequently verified via archival ASAS-SN *V*-band records. In July 2018, my visual observations identified a short-duration fade to magnitude 15.4 that went undetected in concurrent electronic *V*-band data streams, indicating a masking effect likely caused by near-infrared red-leak contamination in the CCD filters. Following a stable phase recorded by continuous CCD *V*-band tracking from 2023 through early 2026, a mid-2026 visual detection caught a full 16-day fading event reaching a minimum magnitude of 15.3. These combined datasets establish that WR 80 undergoes highly variable optical obscuration events.

**Keywords**: stars: individual (WR 80) – stars: Wolf-Rayet – stars: variables: general – techniques: photometric – circumstellar dust – light curves

## 1. Introduction

Historically catalogued as a stable single star and a continuous infrared dust producer (van der Hucht 2001), the late carbon-sequence WC9d Wolf-Rayet star WR 80 has largely escaped detailed long-term visual monitoring, prompting me to add it to my nightly target list in 2017. According to the International Variable Star Index (VSX; Watson et al. 2006), the star is a highly reddened object with an optical colour of $B$-$V$ = +1.78 and a near-infrared colour of $J$-$K$ = +3.15, indicating an extreme environment of structured circumstellar carbon dust production (Zubko et al. 1992). Due to this severe redness, the bulk of its optical light is heavily shifted toward longer wavelengths, creating a unique challenge for standardised filter bands and automated electronic sensors. While previous studies have documented fading behaviours in similar Wolf-Rayet stars (Veen et al. 1998; Williams 2014), this paper presents a visual and digital observing record establishing that WR 80 is not a static object, but instead undergoes dramatic, short-duration optical fading events.

## 2. Observations

All visual observations were performed from the Tetoora Road Observatory using a 22-inch (0.56m) telescope at a magnification of 270× paired with an 8mm Tele Vue Ethos eyepiece. Under optimal observing conditions, this specific optical configuration consistently achieves a visual pinpoint stellar limiting magnitude of 17.3.

Monitoring of WR 80 began in January 2017, with the star initially found at its documented stable optical baseline at magnitude 14.5. The first signs of unusual activity appeared in March 2017, when I detected a fade to magnitude 15.2 lasting 15 days. Another fade occurred 8 days later in early April, again to magnitude 15.2 over a period of 17 days before returning to its baseline of magnitude 14.5, suggesting this behaviour was due to dust events (Williams 2014).

On 26 April 2017 (JD 2457870), the star underwent a massive plunge from magnitude 14.6 to a minimum magnitude of 15.8 over a single night and remained at this deep minimum state the following night. I observed and verified this minimum with high confidence by making a direct comparison against a magnitude 15.8 reference star positioned roughly 60 arcseconds away within my high-magnification field of view. Recognising the significance of this fade, I requested urgent confirmation via a CCD observation. Follow-up CCD *V*-band observations were successful and captured the star during its subsequent rise from magnitude 14.9 *V* back to the varying baseline of magnitude 14.5 *V* to 14.6 *V*, independently validating my visual detection of this rapid recovery phase. The total duration of this deep fade from start to finish was only 6 days (Fig. 1).

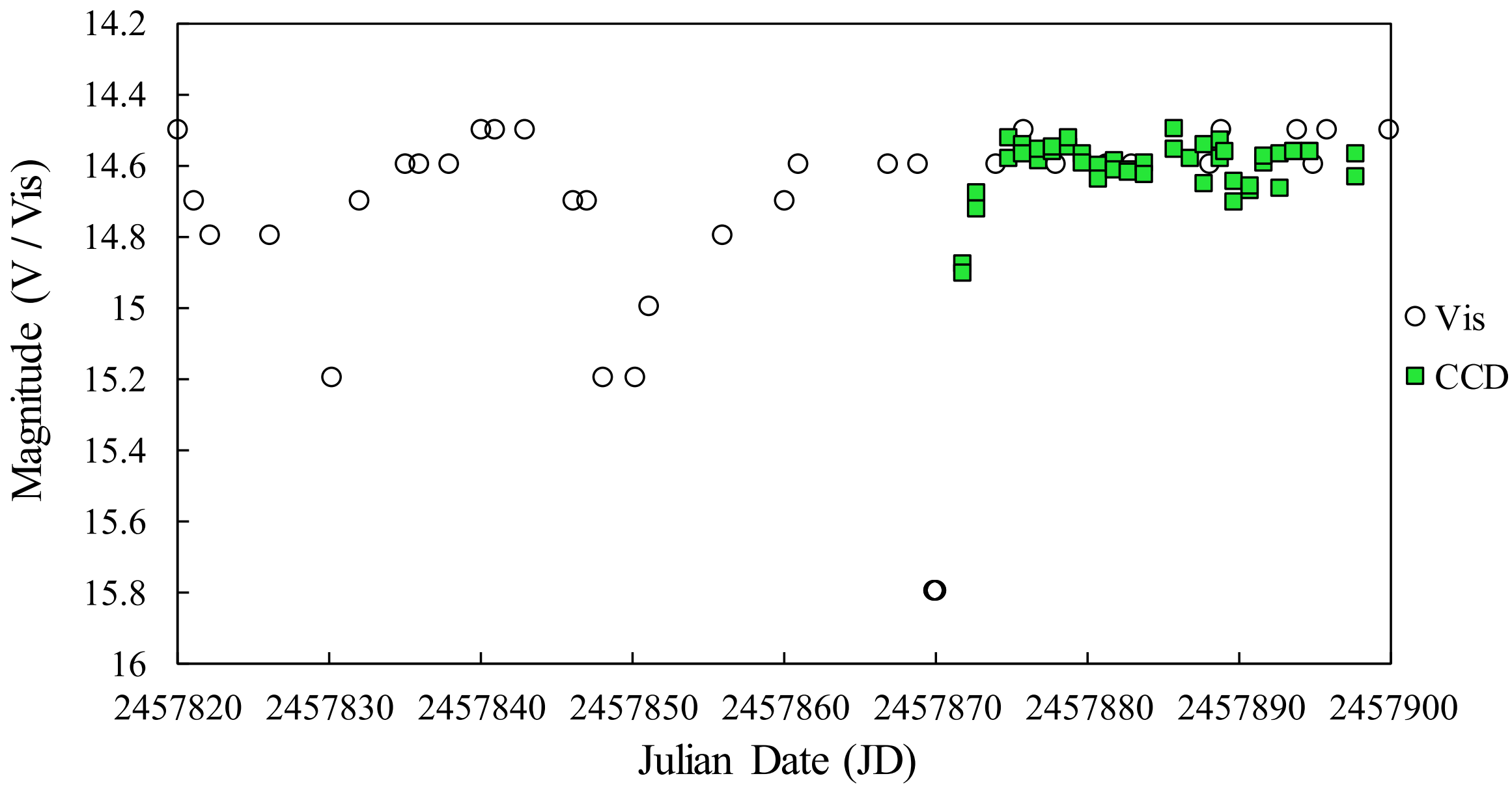

**Figure 1.** Long-term light curve of WR 80 showing early transient dust activity in 2017. The plot documents the initial magnitude baseline, multiple short-duration dips to magnitude 15.2, and the deep visual discovery plunge to a minimum magnitude of 15.8 (open circles). The subsequent rapid 6-day recovery phase back to maximum light and independent confirmation by follow-up CCD *V*-band data (green squares) is shown. WR 80 also shows minor fluctuations along its baseline between magnitudes 14.5 and 14.6.

My monitoring of WR 80 continued, and I detected another rapid fade from magnitude 14.7 in July 2018. I observed the minimum point down to magnitude 15.4 on JD 2458314 over a two-night window. Notably, simultaneous electronic data failed to detect this brief event. This second fade confirmed that these fadings were recurring events, rather than the single isolated fading event in 2017.

Visual observations and CCD *V*-band imaging continued through 2020 with no further fading events detected. No data were obtained for the subsequent three years until CCD *V*-band observations resumed in 2023 and continued through early 2026, recording a comparatively stable baseline of magnitude 14.5 *V* with no major fading events during this timeframe. Visual observations of WR 80 were recommenced in April 2026. The motivation was to detect another fade similar to the events of 2017 and 2018, as these dynamic drops remained undocumented in the literature.

## 3. Archival Confirmation

Following the resumption of my monitoring programme, I examined the historical records of the All-Sky Automated Survey for Supernovae (ASAS-SN; Kochanek et al. 2017) to seek independent verification of these past fading events. The archival ASAS-SN data obtained with the *V*-band filter revealed a distinct drop to magnitude 15.1 *V* in 2017. Crucially, this coincided exactly with JD 2457870, the date of my discovery of the deep visual fade on 26 April 2017. Both my visual estimates and the ASAS-SN *V*-filter datasets independently recorded a rapid, 6-day total duration for this event, providing definite cross-verification of this event.

Looking further back in the ASAS-SN archive, another distinct fading event was detected in mid-2016, to a minimum magnitude of 15.0 *V* and spanned a total duration of 13 days. The time interval between this 2016 event and the deep 2017 event was 390 days.

## 4. The 2018 Anomaly and Filter Transitions

During 2018, ASAS-SN was transitioning its primary monitoring operations from the traditional *V*-filter to the bluer *g*-filter. At the time of the July 2018 fade, ASAS-SN was using the *V*-filter and recorded a flat baseline of magnitude 14.5 *V*, which matched concurrent ground-based CCD data of magnitude 14.5 *V*. In contrast, my visual observations recorded a clear, short-duration drop over two consecutive nights from magnitude 14.7 to magnitude 15.3 on JD 2458313, and a minimum magnitude of 15.4 on JD 2458314, coinciding directly with the night of the 14.5 *V* measurements. The visual observations perfectly matched the magnitude 14.5 *V* electronic baseline datasets both before and after this fading event, confirming that the digital sensors and my eye were calibrated under normal seeing conditions.

The comparison star I used over these two nights was a magnitude 15.2 star located just 1 arcminute away, ensuring a reliable baseline. Remarkably, on the night of the deepest visual minimum (JD 2458314), when I recorded a fade to magnitude 15.4, a concurrent ground-based CCD observation was obtained within just 39 minutes of my visual estimate. Yet, CCD *V*-band data recorded a baseline magnitude of 14.52 *V*. Because WR 80 is a highly reddened object, filter transmission profiles and near-infrared red-leak characteristics can strongly influence the electronic *V*-band data. This red leak likely shielded the rapid optical dip from the CCD sensors.

At the same time, my eye is naturally blind to these near-infrared wavelengths, allowing me to resolve the true visual minimum shown in Fig. 2. This direct overlay highlights how differently electronic and visual observations respond to highly reddened environments. Further long-term monitoring is required before dismissing this event as an anomaly.

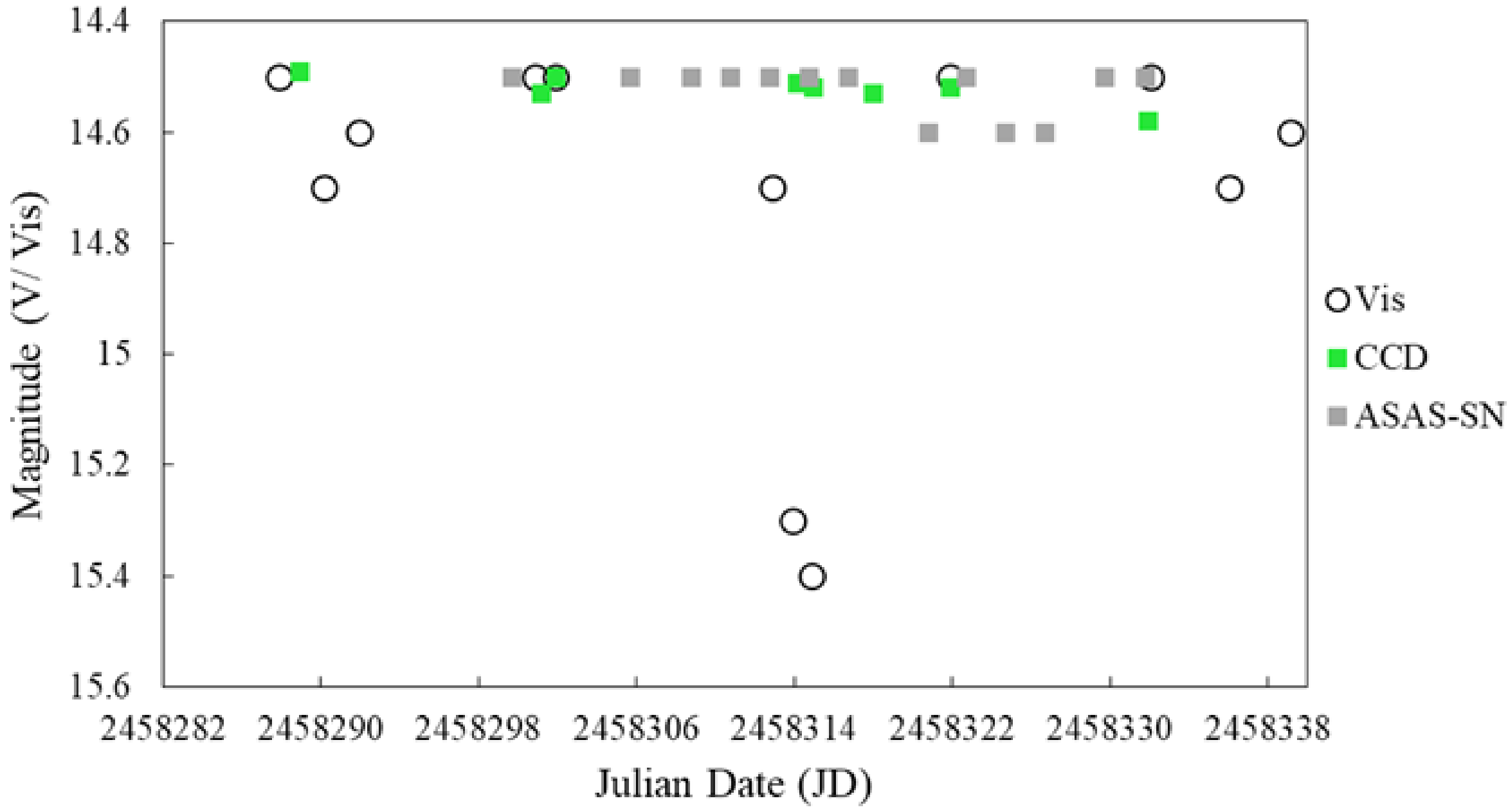

**Figure 2.** Multi-dataset light curve of WR 80 during the short-duration July 2018 event. The plot highlights the difference between the visual observations (open circles) and the electronic *V*-band data. While the visual data display a profound, rapid, two-night plunge that faded to a minimum magnitude of 15.4 on JD 2458314.91, the CCD *V*-band data (green squares) and archival ASAS-SN *V*-filter data (grey squares) remain stable at a baseline of magnitude 14.5 *V*.

Following this event, the All-Sky Automated Survey for Supernovae (ASAS-SN) shifted its operations exclusively to the *g*-filter. This bandpass shifts WR 80 to a significantly fainter baseline, varying from magnitude 15.4 *g* to 15.7 *g*. This transition limits the effectiveness of subsequent automated *g*-filter data for detecting these specific, red-sensitive optical dust fades, which was highly evident during the 2026 fade.

## 5. The 2026 Fading

After an extended quiet phase lasting more than three years, the long-term stable baseline of WR 80 abruptly ended in mid-2026. My visual monitoring programme recommenced in April 2026. The star sat comfortably around its regular baseline, varying between magnitude 14.5 and 14.6. By early June, I detected a major new dust event with the star at magnitude 14.6 up to JD 2461193 before entering a rapid ingress phase. Over a period of approximately 7 days, WR 80 slowly faded 0.7 magnitudes, hitting a well-defined minimum floor of magnitude 15.3 on JD 2461200.86 through JD 2461200.90. Following this minimum hold, I observed the star as it entered a steady egress climb. The recovery phase progressed through magnitudes 15.1, 14.9, and 14.7 before successfully returning to a baseline of magnitude 14.6 by JD 2461210. The total duration of this entire fading event spanned approximately 16 days (Fig. 3).

The comparison star sequence and chart I used for WR 80 have remained identical from 2017–2026, ensuring strict internal consistency across a decade of observations. Data from the ASAS-SN pipeline showed a median *g*-band magnitude of 15.6 *g* before, during and after this same fading period due to the bandpass filter shift, and no ground-based CCD *V*-band data were obtained during this interval. Ongoing long-term monitoring and high-cadence data are required to characterise these fading profiles further and to document whether these events exhibit any long-term, predictable periodicity.

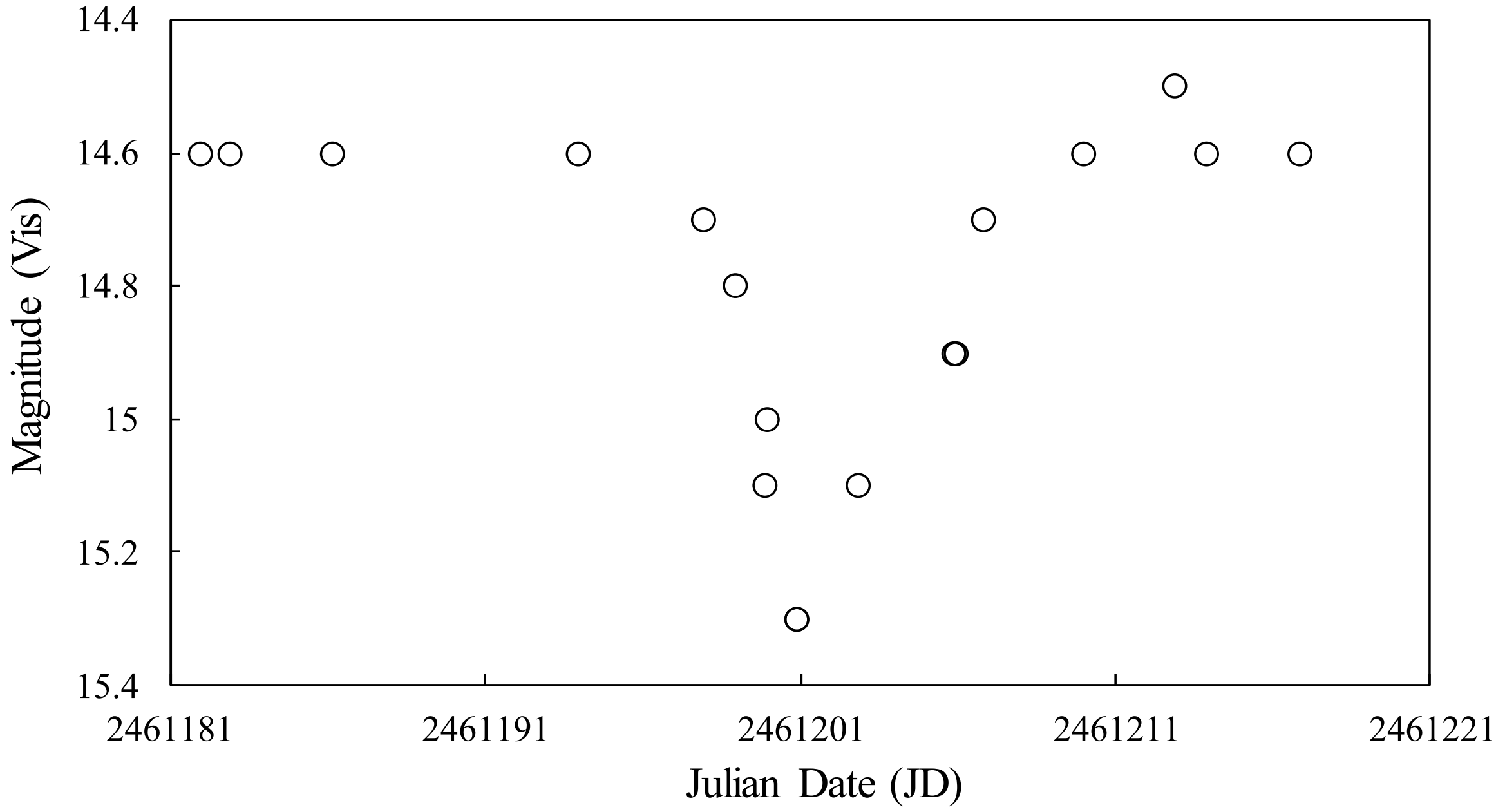

**Figure 3.** High-cadence, well-defined visual light curve capturing the June 2026 dust-fading event of WR 80. The plot illustrates a stable prolonged baseline at magnitude 14.6, a rapid 7-day ingress phase, a floor minimum at magnitude 15.3 centred around JD 2461200, and a structured egress recovery spanning 16 days.

## 6. Conclusion

Consistent monitoring from 2017–2026 has established that the WC9d Wolf-Rayet star WR 80 acts as a highly dynamic, variable dust producer. The first historically recorded fading event occurred in 2017, plunging to a visual magnitude of 15.8. Archival ASAS-SN *V*-band data independently verified the activity, capturing a minimum of 15.1 *V*. These observations confirm that the star is capable of an extremely rapid 1.3-magnitude plunge over a single night and a rapid recovery period lasting just 6 days.

While long-term electronic and automated survey trends can show the star as a relatively stable quiescent object, my visual observations reveal a profoundly different behaviour. The recent 2026 event confirms that these dust-fading events are recurring transient events that emerge unpredictably after years of apparent inactivity. These findings demonstrate that continued monitoring from dedicated visual observers using high-magnification instruments remains absolutely vital to modern stellar astrophysics. By maintaining continuous ground-based monitoring, an experienced visual observer can bypass electronic sensor limitations, such as near-infrared red-leaks, and filter limitations to isolate real-time optical dust-fading events as they occur. Continuous high-cadence monitoring and multi-wavelength data collection are essential to characterise the potential predictability of this variable dust production, and to determine how the specific amplitude of a fade dictates whether automated electronic surveys can detect these events or remain completely blinded by them.

## Acknowledgments

My sincere thanks to Josch Hambsch for providing critical CCD confirmation data during the 2017 egress. This research has made use of the data archive provided by the All-Sky Automated Survey for Supernovae (ASAS-SN). In addition, I gratefully acknowledge the American Association of Variable Star Observers (AAVSO) and the use of data from the AAVSO International Database contributed by observers worldwide.